\documentclass{article}

\usepackage{PRIMEarxiv}
\usepackage[utf8]{inputenc} 
\usepackage[T1]{fontenc}    
\usepackage{hyperref}       
\usepackage{url}            
\usepackage{booktabs}       
\usepackage{amsfonts}       
\usepackage{fancyhdr}       
\usepackage{graphicx}       
\usepackage{subcaption}
\title{Impact of AI-tooling on the Engineering workspace
}

\author{
  Lena Chretien \\
  Jellyfish.co \\
  Boston, MA\\
  \texttt{lena@jellyfish.co} \\
   \And
  Nikolas Albarran \\
  Jellyfish.co \\
  Boston, MA\\
  \texttt{nik@jellyfish.co} \\
}

\begin{document}
\maketitle

\begin{abstract}
To understand the impacts of AI-driven coding tools on engineers' workflow and work environment, we utilize the Jellyfish platform to analyze indicators of change. Key indicators are derived from Allocations, Coding Fraction vs. PR Fraction, Lifecycle Phases, Cycle Time, Jira ticket size, PR pickup time, PR comments, PR comment count, interactions, and coding languages.
Significant changes were observed in coding time fractions among Copilot users, with an average decrease of 3\% with individual decreases as large as 15\%. 
Ticket sizes decreased by an average of 16\% across four companies, accompanied by an 8\% decrease in cycle times, whereas the control group showed no change. Additionally, the PR process evolved with Copilot usage, featuring longer and more comprehensive comments, despite the weekly number of PRs reviewed remaining constant.
Not all hypothesized changes were observed across all participating companies. However, some companies experienced a decrease in PR pickup times by up to 33\%, indicating reduced workflow bottlenecks, and one company experienced a shift of up to 17\% of effort from maintenance and support work towards product growth initiatives.\\
This study is the first to utilize data from more than one company and goes beyond simple productivity and satisfaction measures, considering real-world engineering settings instead. By doing so, we highlight that some companies seem to benefit more than others from the use of Copilot and that changes can be subtle when investigating aggregates rather than specific aspects of engineering work and workflows - something that will be further investigated in the future. 
\end{abstract}

\section*{Introduction}
The emergence of artificial intelligence (AI)-enabled coding tools represents a paradigm shift in contemporary software development and engineering workflows. Prominent among these innovations is GitHub Copilot, an AI-driven code completion tool collaboratively developed by OpenAI and GitHub. Leveraging the capabilities of the Codex model, which is trained on extensive corpora of publicly accessible code, GitHub Copilot aids developers by recommending code snippets and even complete functions derived from natural language descriptions.

A substantial body of research has addressed the robustness and security implications of this tool \cite{mastripaolo, zhang, dakhel, jaworski, nguyen}. Concurrently, several empirical investigations have evaluated the professional utility and productivity impacts of GitHub Copilot \cite{mozannar, vaithilingam, peng}. Collectively, these studies underscore the potential benefits that such tools offer to software developers and engineers, particularly when employed within appropriate contexts. Nonetheless, they also caution against potential pitfalls— specifically, the risk that developers lacking sufficient contextual understanding or coding expertise may inadvertently accept flawed code suggestions.

Most studies to understand the impact of Copilot on engineering satisfaction were conducted through surveys which highlighted the overall satisfaction of engineers when able to use Copilot \cite{github, bench}, with some increases up to 50\%. 

One recent study was conducted to determine the enhancement in coding velocity and the potential significant productivity advancements (\cite{peng}). This investigation, conducted within a controlled setting, involved 95 freelance engineers tasked with a specific coding challenge. Half of the engineers received assistance from GitHub Copilot while the other half did not. The results revealed a 55\% reduction in task completion time when the AI tool was utilized.

Despite these findings, a recurring theme across these studies, as well as additional small pilot studies within smaller companies, is the prevalent focus on productivity metrics, while neglecting a holistic examination of daily engineering workflows, behavioral shifts, and broader engineering dynamics. This together with the lack of data driven insights from a large number of Copilot users are the two main factors that are  currently limiting our understanding of impacts of AI-driven coding tools on the engineering workspace.  

This paper seeks to bridge this gap by examining the impacts of the AI tool GitHub Copilot on engineers' workflows, productivity, and daily interactions. Rather than relying on surveys, we leverage empirical data sourced from the Jellyfish platform, encompassing GitHub, Jira, allocations, metrics, and more. This dataset comprises real-world activities of 133 engineers from four different companies who have integrated GitHub Copilot into their operations over periods ranging from one to eight months. We conduct a comprehensive before-and-after analysis of these engineers' work and behavior, supplemented by comparative evaluations with a control group of engineers who did not use GitHub Copilot. Hence, we provide insights into more than productivity metrics, considering the real-world applications of the tool in a day-to-day setting. 

\section*{Hypothesis}
\label{hypo}
\begin{figure}
    \includegraphics[scale=0.25]{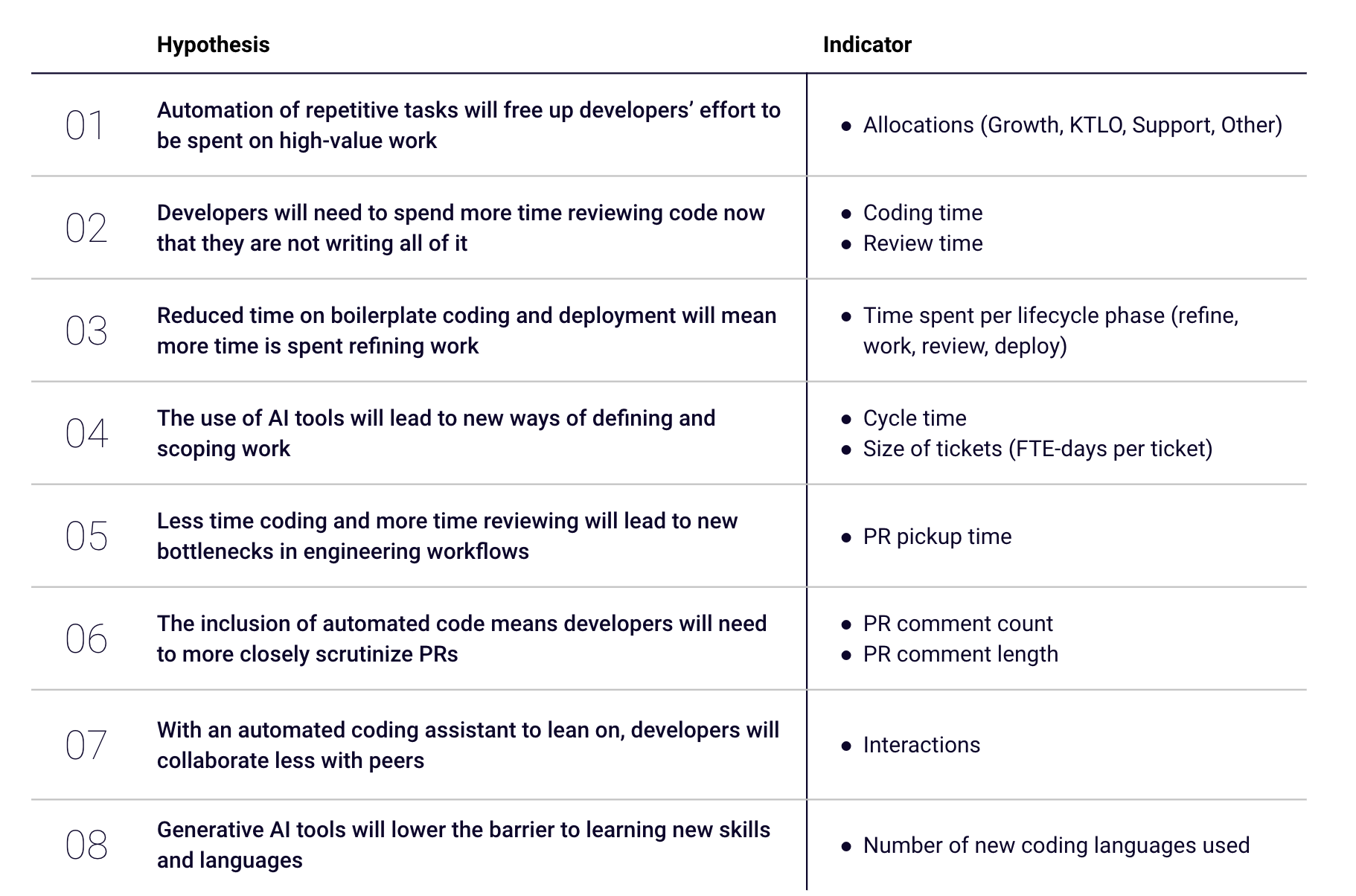}
    \caption{Hypotheses of changes observed due to usage of Copilot and indicators that can be used to investigate these changes.}
    \label{hypothesis_table}
\end{figure}

This study proposes several transformative changes in the engineering workflow, engineering operations, and strategic alignment within software organizations due to the integration of the AI-based coding assistant, Copilot. Specifically, we formulate eight hypotheses delineated in Figure \ref{hypothesis_table}. 

Each hypothesis will be evaluated using a unique set of indicators derived from the extensive Jellyfish dataset (Figure \ref{hypothesis_table}), which includes GitHub metadata on commits, pull requests, reviews, Jira Metadata on tickets, statuses, assignee, comments, as well as allocations, time, and effort spent on tasks.

For instance, Hypothesis 1 suggests that a decrease in time spent on repetitive tasks will manifest as an increase in allocations towards high-value 'Growth' work. Conversely, this should logically entail a decrease in 'Other' categories such as KTLO, Support, etc. \\
Repetitive tasks and coding that still takes place will be done with the assistance of an AI coding tool and so (Hypothesis 2) coding will become quicker, while the amount of review for each piece of work will increase to ensure correctness of AI-implemented code (or partial code). \\
As Copilot is built to enhance coding and deployment efficiencies (Hypothesis 3) it is anticipated that more resources will be directed towards the creative tasks of refinement. This shift will be evident in the lifecycle phases of tickets, with the noted accompanying increase in review activities.\\
Furthermore, as coding becomes a less taxing component of a developer's workload, the nature of task definition is expected to evolve (Hypothesis 4). This evolution will manifest in smaller work units (effort per ticket) and reduced cycle times per ticket. \\
With the acceleration of coding processes and a heightened focus on review activities, new procedural bottlenecks are anticipated (Hypothesis 5) as would be observable in extended PR Pickup times. \\
The emphasis on more rigorous review processes is also expected to result in more thorough and detailed PRs and PR comments (Hypothesis 6) as evident in an increased number and length of comments. \\ 
Regarding the day-to-day interactions of engineers, the integration of an automated coding assistant is predicted to result in reduced collaboration (Hypothesis 7), as reflected in the decreased number of interactions such as comments, edits on peers' work, etc. Concurrently, the removal of barriers to acquiring new skills and utilizing new coding languages (Hypothesis 8), will be evidenced by an increase in the diversity of coding languages employed. 

Some of these hypotheses can be thought of as "leading impacts" - operational metrics like coding time, cycle time, collaboration, review times, and others are "lagging impacts" such as bottlenecks, effort allocations, and new skills. In other words, for some areas, impacts might be stronger and observed sooner than in others.  

\section*{Study setup}
To better understand the evolution of the engineering workspace, our investigation explores the impact of Copilot usage in everyday environments, rather than a controlled or semi-controlled experiment setup as referenced in prior studies \cite{peng, bench}. Our study was conducted in collaboration with four distinct companies from our customer base, involving 133 engineers who utilized Copilot in conjunction with the Jellyfish platform. The study was conducted using a retrospective design, ensuring that the engineers' behavior and workflows remained unaffected by the awareness of the study taking place. For each engineer, we obtained the initial date of Copilot usage, along with comprehensive data from the Jellyfish platform, including Jira tickets, GitHub activity, and Allocations. With this, we investigate changes in work behavior and workflows for the period before Copilot adoption and the period of Copilot usage. 

\begin{figure}\includegraphics[scale=.65]{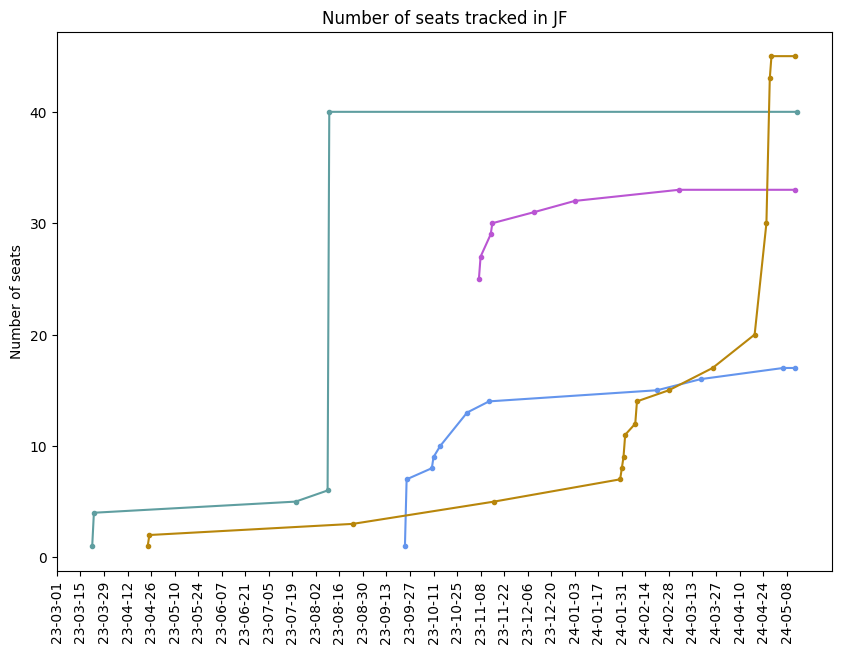}
\centering
    \caption {Copilot seats over time, by company}
    \label{seats}
\end{figure}
The companies in this study integrated Copilot into their operations as early as March 2023. Initially, there was a limited number of active seats, which subsequently expanded to between 17 to 45 seats per company by May 2024 (Figure \ref{seats}). 

Our methodology for this study employed a comparative analysis for each Copilot user's performance, evaluating the three months prior to and subsequent to the adoption of the AI tool. Furthermore, in order to isolate the effects of the Copilot adoption, we conducted a comparative assessment between the Copilot-user group and a control group. This control group comprised 750 engineers from the same companies who, while active on the Jellyfish platform, did not use Copilot. 

We investigated the hypothesis mentioned in Figure \ref{hypothesis_table} by examining indicators for each, with data from the Jellyfish platform. The data and indicators are detailed in the subsequent section.   
 
\subsection*{Data} 
We investigate indicators derived from Jellyfish data, including: Allocations, Coding fraction vs PR fraction, Lifecycle Phases, Cycle time, Size of Jira tickets, PR Pickup time, PR comments, PR comment count, Interactions, and coding languages. 

\underline{Allocations:}\\
This metric quantifies how a software engineer's efforts are distributed across different areas of work: KTLO (Keeping the lights on), Growth, Support, and Other \cite{allocations}. The specific nature of these categories is customized to the needs of each customer and captured and measured via Jellyfish. Allocations are quantified in full-time engineers (FTEs), a standard unit that mitigates inconsistencies found in other measures such as story points or issue counts \cite{story_points}. 

\underline{Coding fraction vs. PR fraction, and Cycle time:}\\ 
By integrating insights from Jira issues with GitHub activities, we avoid merely calculating cycle time in isolation. Instead, we can get a more nuanced view and consider work only if it involved coding (at least one commit was associated with a Jira ticket), as well as split cycle time into a) time spent coding, vs b) the time spent under Review. To do this we capture the first time instance of when a ticket was moved into the Indeterminate phase (this is the phase that signals active work on a ticket). The time spent coding is the time until the first PR was opened for this work. The time in review, on the other hand, is the time from the first PR to the merging of this PR. 
Each phase's duration (coding and review) is expressed as a fraction of total work length, standardizing variations of ticket size and work segmentation. \\ 
The total length of the ticket - that is the coding time plus the review time - is referred to here as cycle time.

\underline{PR Pickup time, number of PRs, and PR comments:}\\
The duration a ticket remains in the PR review phase depends not only on the assigned engineer but also on the timeliness of PR reviews. 
PR Pickup time is defined as the interval from PR opening to the first review. This metric is captured by keeping track of the times of PRs and any associated activity such as comments, reviews, etc. 
Furthermore, by analyzing data on the number of pull requests (PRs) created and reviewed by each engineer, as well as the comments made on these PRs, we are able to quantitatively assess individual PR activity. Additionally, the aggregate length of comments on each PR provides a metric for evaluating the thoroughness and comprehensiveness of the reviews conducted.

\underline{Size of Jira tickets:}\\
In addition to cycle time per ticket - a good indicator of how large a piece of work might have been - we also explore the effort that was associated with each ticket. This is measured by accumulating actual work signals (e.g. comments, edits, commits, PRs, etc.). 
We do this to eliminate any outside factors that might contribute to extended cycle times, such as bottlenecks, or blocks because of dependencies in work. In particular, cycle time could be large while effort was not if an engineer moved a ticket into the Indeterminate status in Jira and started work but was consequently blocked or pulled into different work which interrupted their focus. In this case, the clock for cycle time would keep counting up, while effort - defined as actual signals from Jira or GitHub activity - would not be detected and remain low. 

\underline{Lifecycle Phases:}\\
The lifecycle explorer is a feature within the Jellyfish platform designed to provide insights into the software development process by breaking work down into distinct phases. The phases track the progress of the work from definition to completion, helping engineering leaders visualize and understand how time and resources are allocated throughout the development cycle. There are four phases in the lifecycle of work: Refinement, Work, Review, and Deployment. 

\underline{Interactions:} \\
Identifying connections between engineers is accomplished through counting upstream and downstream interactions through the count of comments and edits made or received between team members on Confluence pages, PRs, and Jira issues. This allows insights into the collaboration or isolation of engineers, and provides a better understanding of team dynamics and knowledge sharing. 

\underline{Coding Languages:}\\
Although Jellyfish does not directly access the code committed to GitHub, the extension on files that were worked on serves as proxies for understanding the variety of coding languages employed over time. 

This data-centric approach allows us to rigorously analyse the real-world impacts of AI tools on software engineering workflows and productivity.

\section*{Results}
\begin{figure}
\centering 
    \begin{subfigure}{0.7\linewidth}
        \includegraphics[scale = 0.5, width=\linewidth]{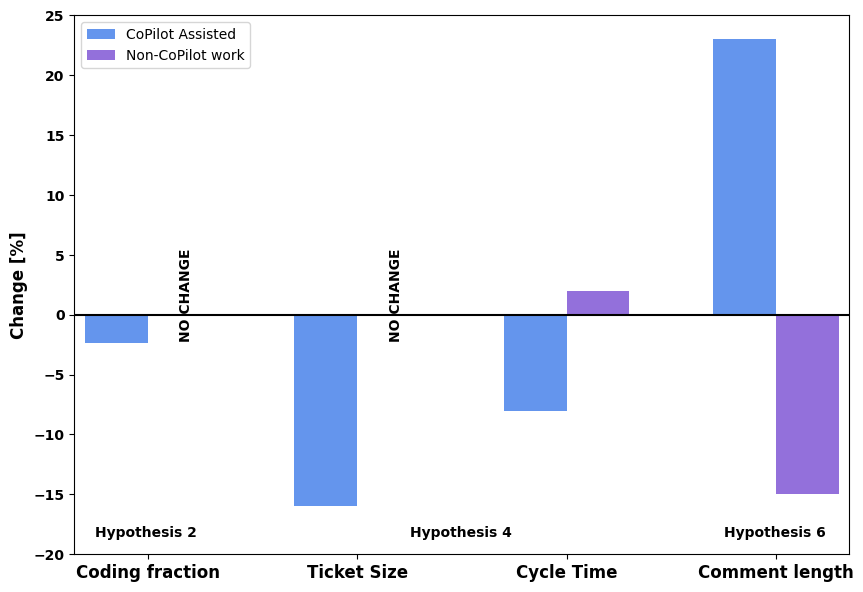}
        \caption{Hypothesis with changes}
        \label{changes}
    \end{subfigure}

    \begin{subfigure}{0.7\linewidth}
        \includegraphics[scale = 0.5, width=\linewidth]{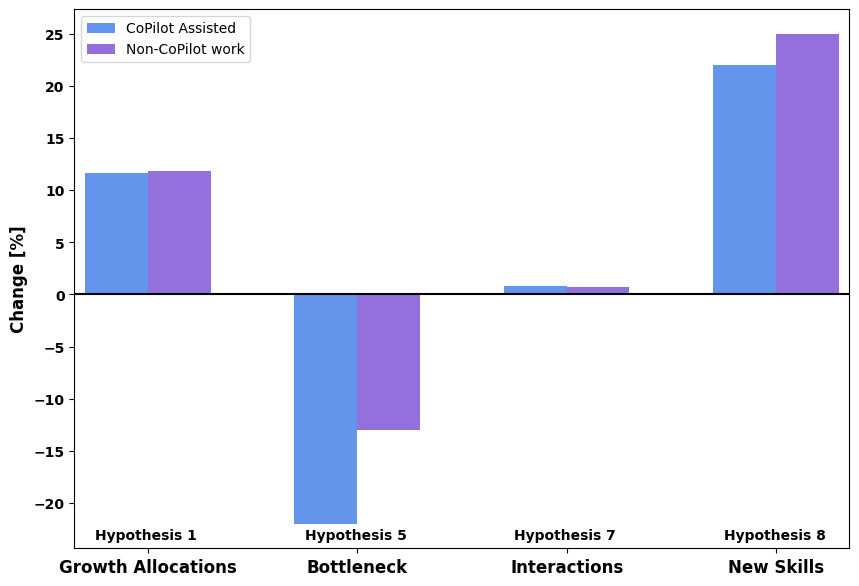}
        \caption{Hypothesis without changes}
        \label{non-changes}
    \end{subfigure}
    \caption{Relative changes [\%] in the before and after of Copilot users (blue) vs. non-Copilot users (purple) for Hypothesis 2, 4, and 6 (a) and Hypothesis 1, 5, 7, and 8 (b) (Figure \ref{hypothesis_table})}
\end{figure}

In our comparative analysis of Copilot users and non-users, significant differences were observed particularly in Hypothesis 2, 4, and 6 (Figure \ref{changes}). Conversely, the indicators related to the other hypotheses did not exhibit significant changes when compared to those observed in the control group (Figure \ref{non-changes}). 

Significant changes were observed in coding time fractions. Specifically, the fractions that Copilot users spent coding per task ranged from a decrease of 15\% to an increase of 1\%, with an average of 3\% decrease. In contrast, the control group showed changes ranging from a decrease of 6\% to an increase of 3\%, with no average change. This difference is statistically significant with a p-value of 0.01, indicating that the two groups originate from distinct populations. The average coding fraction per ticket remained at 34.25\% ($\pm$ 8\% and 9\% for the before and after timeframe, respectively) for the control group but decreased from 31.5 $\pm$ 6\% to 30 $\pm$ 2\%. 

Work definition (Hypothesis 4) showed notable changes as well, particularly in how work was scoped. Knowing that a Jira ticket - a piece of work - can vary in size, the FTE count from the Jellyfish platform provides a reliable measure of the actual work associated with each ticket. Using this data, we observed that post-Copilot adoption, ticket size decreased across all four companies with an average reduction of 16\%. This was accompanied by a nearly 8\% decrease in cycle times, while the control group saw no change for the former, and a slight increase for the latter.   
In other words, work was not just defined in smaller chunks - i.e. less work associated with each Jira ticket - but this work was also completed more quickly, as evidence by the decreasing cycle time. 
The decrease in ticket size and the subsequent scoping of work into smaller chunks align with current CI/CD best practices. For some companies, we observe coding ticket sizes decrease by as much as 33\%.

Additionally, the PR process seemed to evolve with the usage of Copilot, characterized by longer and more comprehensive comments. This was consistent across all four companies, although the weekly number of PRs reviewed remained unchanged. Conversely, the control group experienced a decrease in number and length of comments.   

As mentioned above, while significant changes aligned with Hypotheses 2,4, and 6 for Copilot users, the indicators for other hypotheses did not show any conclusive differences in the change from the before and after periods when compared to the control group. 
However, while not consistently observed across all companies, in some cases PR pickup time decreased up to 33\%, suggesting a reduction in potential bottlenecks in developers' workflows. 
In addition, in some cases a shift of upwards of up to 17\% of effort from KTLO (Keeping the lights on) and support work toward product roadmap and growth work was observed.

\section*{Conclusion}
In our study, three out of the eight hypothesized changes were substantiated by the empirical data, while we saw no impact of Copilot on the other areas of engineering work. This suggests that while Copilot influences certain specific aspects of engineering workflows, its impact on the other dimensions remains comparable to the standard practices of non-users

A nuanced reflection of the day-to-day operations of individual engineers, teams, and entire organizations, sheds light on why the data did not reveal more changes. AI-driven coding tools are built to accelerate coding tasks; however, coding constitutes merely a fraction of an engineer's daily activities. Consequently, the pronounced transformations observed in the controlled experiment conducted by \cite{peng} are mitigated in a typical work environment, where non-coding tasks and workflows, such as reviews, bottlenecks, dependencies, unplanned work, planning, meetings are predominant. This is especially pronounced when aggregating data at the team level - coding tasks become even less prominent, thereby diminishing the observable impacts. Therefore, the influence of AI tooling is more discernible at an individual level than at a team or organizational level, where the multiplicity of non-coding activities dilutes the enhancements provided by such tools.

\begin{figure}                                                
    \includegraphics[scale=0.25]{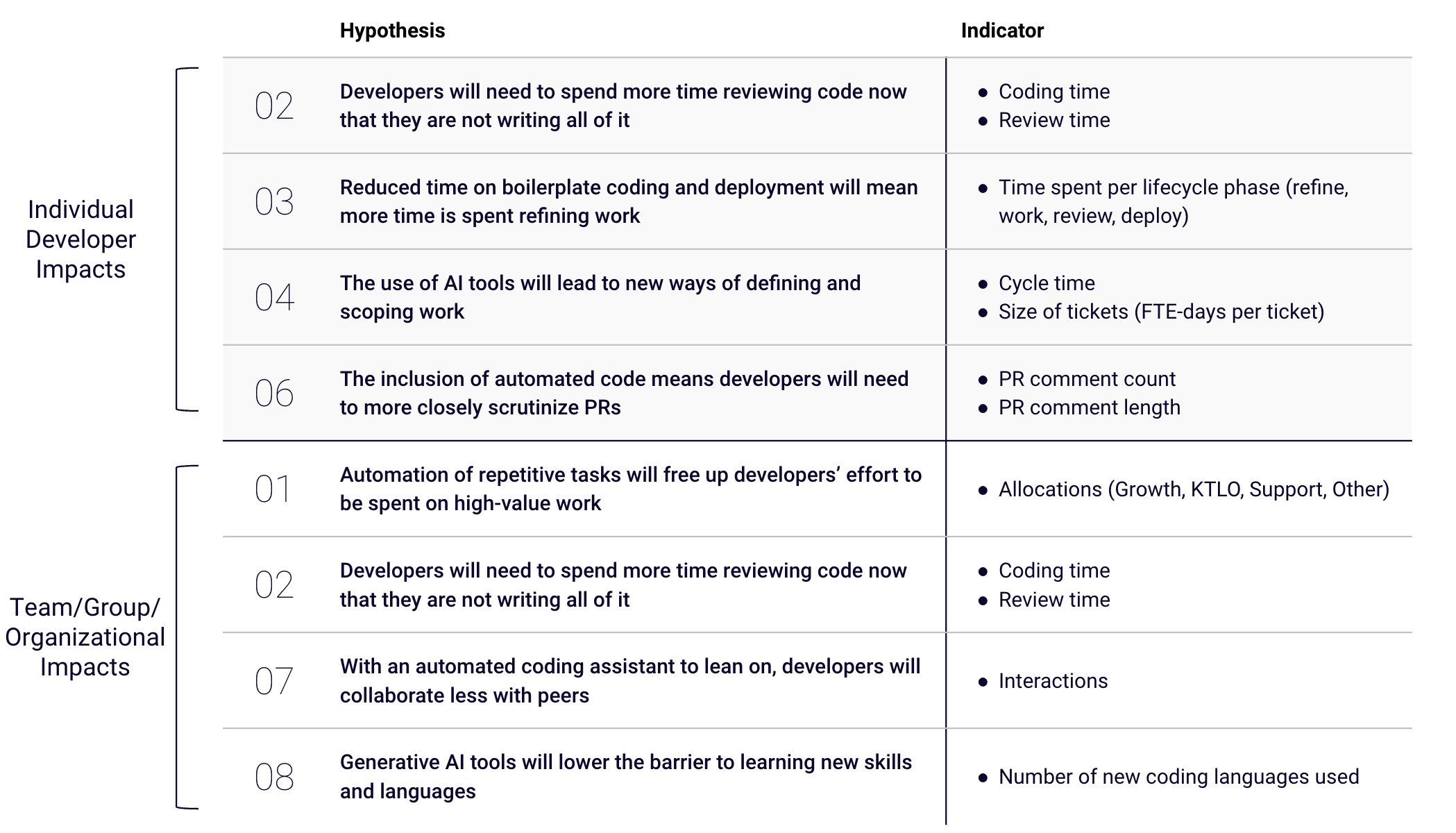}
    \caption{Hypotheses introduced in Figure \ref{hypothesis_table} but sorted by Individual Developer Impacts, and Team/Group/Organizational Impacts}
    \label{hypothesis_table_conc}
\end{figure}

The presented hypotheses can be categorized based on their impact on individual versus aggregated levels, such as multiple individuals and teams (Figure \ref{hypothesis_table_conc}). Processes such as coding versus pull request fractions on each task, task definition, and the review process are primarily individual-centric activities that are likely to be directly impacted by an AI coding tool.

While one might consider "New Skills" in this individual-centric category, we argue that this is not appropriate. An engineer's role within a team or organization typically does not change overnight. Consequently, the type of work (e.g. front-end, back-end, infrastructure, research, etc.) remains relatively constant. As a result, the adoption of new coding languages is expected to occur at similar rates among both Copilot users and non-users. Instead, it might be more pertinent to investigate the onboarding rate of new employees with and without the use of Copilot. It is reasonable to assume that new employees and junior engineers are able to take on more challenging tasks quickly with the help of AI-coding tools. 

Conversely, bottlenecks, Allocations, and interaction are dependent on the team as a whole, the team's processes, and the team's role within the organization. Changes in these areas might be observed if an entire team adopted Copilot; however, influence from other parts of the organization are likely to attenuate these impacts even then. Therefore, it is unsurprising that changes for these hypotheses were not evident, given that only a small fraction of engineers in each company are Copilot users. 

Changes are subtle and even non-existent in many places, which we believe is due to the noise and low resolution of the data. Early findings from a follow-up study using the Copilot API for high-resolution usage data show that Copilot is more heavily used in some areas versus others. We expect to see different levels of impact when aggregating data by e.g. type of work, coding language, seniority of engineer, area of work, etc.  

\bibliographystyle{unsrt}  
\bibliography{Copilot_study_Jellyfish}  

\end{document}